\documentclass[english]{emulateapj}

\usepackage[T1]{fontenc}
\usepackage[latin9]{inputenc}
\usepackage{array}
\usepackage{float}	
\usepackage{amsmath}
\usepackage{color}

\makeatletter

\@ifundefined{textcolor}{}
{%
 \definecolor{BLACK}{gray}{0}
 \definecolor{WHITE}{gray}{1}
 \definecolor{RED}{rgb}{1,0,0}
 \definecolor{GREEN}{rgb}{0,1,0}
 \definecolor{BLUE}{rgb}{0,0,1}
 \definecolor{CYAN}{cmyk}{1,0,0,0}
 \definecolor{MAGENTA}{cmyk}{0,1,0,0}
 \definecolor{YELLOW}{cmyk}{0,0,1,0}
 }

\makeatother

\usepackage[normalem]{ulem}  

\usepackage{babel}

\usepackage{natbib}

\shorttitle{A new crystalline phase in magnetar crusts}
\shortauthors{Bedaque et al.}

\begin{document}

\title{A new crystalline phase in magnetar crusts}
\author{ Paulo F. Bedaque }
\author{Simin Mahmoodifar}
\author{Nathan Ng}
\author{Srimoyee Sen}
\affil{Department of Physics, University of Maryland College
Park, MD 20742, USA}

\begin{abstract}
We show that ions at the low densities and high magnetic fields relevant to the outer crust of magnetars form a novel crystalline phase where ions are strongly coupled along the magnetic field and loosely coupled in the transverse direction. The underlying cause is the anisotropic screening of the Coulomb force by electrons in the presence of a strongly quantizing magnetic field which leads to Friedel oscillations in the ion-ion potential. In particular, the  Friedel oscillations are much longer-ranged in the direction of the magnetic field than is the case in the absence of magnetic fields, a factor that has been neglected in previous studies. These ``Friedel crystals'' have very anisotropic elastic moduli, with potentially interesting implications for the Quasi-periodic Oscillations seen in the X-ray flux of magnetars during their giant flares. We find the minimum energy configuration of ions taking into account these anisotropic effects and find that, depending on the density, temperature and magnetic field strength, different lattice structures (fcc, hcp or bcc oriented in different ways in relation to the magnetic field)  are favored.
  \end{abstract}


\section{Introduction}
The observations of quasiperiodic oscillations (QPOs) in the X-ray flux of magnetars (highly magnetized neutron stars with $B \gtrsim 10^{14}$ G) \citep{1998ApJ...498L..45D, 2005ApJ...628L..53I, 2005ApJ...632L.111S, 2006ApJ...653..593S, 2006ApJ...637L.117W, 2006csxs.book..547W}, which have been linked to global torsional vibrations within the star's crust,  provide a promising new probe of a neutron star's internal composition and structure. Up to now these oscillation have been modeled assuming the crust to be an isotropic elastic material  \citep{1988ApJ...325..725M,1991ApJ...375..679S,1998ApJ...498L..45D,2005ApJ...634L.153P,2007MNRAS.374..256S} but the strong magnetic field present at the crust may change this picture.  It is thus  important to understand the solid state physics of highly magnetized crusts. The structure  of the solid crust is a consequence of the interplay between the ordering influence of the screened Coulomb force between the ions  and the disordering effect of the temperature.
 \cite{2011PhRvC..83b5803S} recently pointed out that the usual screening of the Coulomb force between ions changes substantially by strong magnetic fields. The main change for our considerations, is that the  oscillatory part  of the potential (Friedel oscillations) is much longer-ranged than its counterpart  in the no magnetic field case. \cite{2011PhRvC..83b5803S} conjectured that even though the Friedel part of the screened potential is weaker than the Yukawa part at short distances, its long range might have important consequences for the crystal structure.

  More recently \citet{2013PhRvC..88e5801B} studied the effect of the strong magnetic fields on the structure and elastic properties of magnetar crusts taking into account the  anisotropic electronic screening between ions that had been neglected in previous studies  \citep{2013MNRAS.433.2018B,2013A&A...550A..43P,2009PhRvE..80d6405B,Horowitz:2008xr,1991ApJ...383..745L}.
    The picture that emerged in   \citet{2013PhRvC..88e5801B} was that the long-ranged oscillatory force between ions separated by long distances, of the order of thousands of lattice spacings, added coherently and placed the ions into a strongly bound chain aligned along the magnetic field. The ions, in that case, will be separated in the longitudinal direction by the distance $\pi/k_e$ (or a multiple of $\pi/k_e$), where $k_e$ is the electron Fermi momentum in the longitudinal direction. Crystals with that property will be denoted as ``Friedel crystals'' in this paper.  The interactions  among the chains are much weaker and determine the crystal structure in the transverse direction. By the use of a simple model, \citet{2013PhRvC..88e5801B}  delineated the regions of parameter space where this picture might emerge and estimated the elastic constants (the bulk and shear moduli) that are dominated by the longitudinal structure of the lattice.  In fact, it was shown  that due to the coherent effect of thousands of ions some elastic moduli  are significantly larger than those of a usual bcc Coulomb crystal at comparable densities and no magnetic field.  Since the frequencies of the crustal torsional modes are a function of the shear modulus of the crust, this has potentially interesting implications for magnetar QPOs.
The simple model in \citet{2013PhRvC..88e5801B}, however, cannot address the transverse structure of the lattice that determines the remaining elastic moduli.  In order to be able to make a full analysis of the elastic constants in all directions, including the ones that depend on the direction perpendicular to the magnetic field, one needs to determine the exact crystal structure, including the transverse direction.

  The goal of this paper is  twofold. First, we would like to verify, strengthen and generalize the predictions of the formation of the strongly coupled filaments of ions along the magnetic field (Friedel crystals) in the parameter space relevant to the magnetars which were obtained in  \citet{2013PhRvC..88e5801B} using many simplifying assumptions. Secondly, we would like to determine the exact crystal structure of ions including the transverse direction.

\section{Anisotropic electronic screening and Friedel potential}

In the outer crust of magnetars, the magnetic field strongly quantizes the motion of electrons perpendicular to the field into Landau orbitals and most of the  electrons occupy only the lowest Landau level. Let us consider a uniform magnetic field in the z-direction, $\mathbf{B}=B \hat{z}$. The relation between the electron Fermi momentum, $k_e$, in the z-direction and the electron density, $n_e$, is
\begin{equation}\label{eq:ne}
n_e = 
\frac{eB}{2\pi^2}k_e,
\end{equation} From Equation~(\ref{eq:ne}) we can find the mass density $\rho$ at which the electrons become relativistic
\begin{equation}
 \rho \approx \frac{A M m eB}{2\pi^2 Z} \approx 0.79\times 10^8 {\rm g/cm}^3 \left(\frac{A_{66}}{Z_{28}} \right)B_{15},
\end{equation} where $B_{15}=B/10^{15}$G, $M$ is the nucleon mass  and we express the mass ($A$) and atomic ($Z$) number of the ions in terms of $A_{66}=A/66$  and $Z_{28}=Z/28$ (chosen because  $^{66}Ni_{28}$ is one of the favored isotopes at these densities).
We are interested in the regime where the field is strongly quantizing and all (or most) electrons are in the lowest Landau level. This corresponds to $T\ll \hbar \omega_c/k_B \simeq 1.3434 \times 10^{11} B_{15}$ K, where $\omega_c=\frac{e B}{m_e c}$ is the electron cyclotron frequency, and the mass density given by
\begin{equation}
\rho \alt \frac{A M}{\sqrt{2}\pi^2 Z}(eB)^{3/2} \approx  5.2 \times 10^8 {\rm g/cm}^3\ \left(   \frac{A_{66}}{Z_{28}}\right)B_{15}^{3/2}.
\end{equation}
Thus, for almost all the parameter space we are interested in, most electrons are in the lowest Landau level and the effect of the magnetic field on electron motion is appreciable. 

The effect of the magnetic field on the electron screening of Coulomb forces was discussed in \cite{2011PhRvC..83b5803S}. The potential between ions at leading order in the fine structure constant $\alpha$ was found to be
\begin{widetext}
\begin{equation}\label{eq:V_ion}
V(r_\perp, z) = Z^2 \alpha
\left[
\frac{   e^{-m_D \sqrt{r_\perp^2+z^2}}  }{\sqrt{r_\perp^2+z^2}}
- \frac{m_D^2 e^{-z/\lambda_T} }{4 z} \frac{\cos(2 k_e z) r_\perp}{{\sqrt{4k_e^2+\frac{m_D^2}{2}\ln(4k_e z)}} }  K_1\left(r_\perp \sqrt{4k_e^2+\frac{m_D^2}{2}\ln(4k_e z)}\right)
\right],
\end{equation}
\end{widetext}
where  $r_\perp$ is the direction perpendicular to $z$, $m_D$ is the Debye mass, where $m_D^2=e^3Bm/(2\pi^2 k_e)$, $K$ is a modified Bessel function of the second kind, and $\lambda_T = 2\pi k_e/(mT)$.  Equation~(\ref{eq:V_ion}) is valid for $z\gg 1/k_e$ and $r_\perp \gg 1/\sqrt{ e B}$. We will require an expression valid for the $r_\perp \ll 1/\sqrt{ e B}$ regime as well. This expression, $V(r_\perp=0,z)$, can be found by going back to a more general form for the potential \citep{2011PhRvC..83b5803S} and performing the appropriate approximations. In this case the Yukawa part remains the same while the Friedel part is given by
\begin{align}
V_F(0, z) 
&\approx
 -\frac{Z^2 \alpha m_D^2}{4 z}  \frac{e^{-z/\lambda_T}  \cos(2 k_e z)}{4 k_e^2+\frac{m_D^2}{2}  \ln(4k_ez)}\nonumber\\
& \times  f\left(\frac{2k_e^2+\frac{m_D^2}{4}\ln(4k_e z)}{eB}\right),
\end{align}
 with
\begin{equation}
f(x) = 1+x e^x E_i(-x),\label{eq:f}
\end{equation}  where $E_i(x)=\int^\infty_x e^{-t}dt/t$ is the exponential integral function. The function $f(x)$ approaches $1$ as $x\rightarrow 0$, and $1/x$ as $x\rightarrow\infty$.
The argument of $f$ is small and therefore $f\approx 1$ for most of the parameter space, except for the $B_{15}\ll 1$, $\rho_8\sim 1$ and a small region where $B_{15} \agt 50 \rho_8$ in which $f\alt 1$. As we will see later, these regions are not important for our purposes as a Friedel crystal will not form there.

We note that the potential  in Equation~(\ref{eq:V_ion}) can be decomposed into an isotropic part describing a shielded Coulomb potential (Yukawa part) and an anisotropic part describing the Friedel oscillations. Equation~(\ref{eq:V_ion}) contains several length scales. The parameter $\lambda_T$,  the length scale over which the  Friedel part of the potential is cut off at finite temperature, is by far the longest one. Numerically it is given by
 \begin{equation}
 \lambda_T = \frac{2\pi k_e}{mT} \approx 1.6\times 10^{-7} cm \left(  \frac{Z_{28}}{A_{66}}    \right) \left( \frac{\rho_8}{B_{15} T_1} \right) ,
 \end{equation} 
 where $T_1=T/$(1keV) and $\rho_8=\frac{\rho}{10^8 g/cm^3}$ , while the other scales are
 \begin{align}
 \frac{1}{m_D}  &\approx   1.3\times 10^{-10}\ cm\  \sqrt{\frac{   Z_{28} \rho_8}   {A_{66}  }  }\frac{1}{B_{15}},\\
 \frac{1}{k_e}   &\approx   3.0\times 10^{-11}\ cm\   \left(  \frac{A_{66}}{Z_{28}}  \right)  \left(  \frac{B_{15}}{\rho_8}  \right) ,\\
 \frac{1}{\sqrt{eB}}  &\approx  8.1 \times 10^{-12}\ cm \ \frac{1}{\sqrt{B_{15}}}.
 \end{align} 
 This disparity in length scales means that the Friedel term is very long ranged compared to either the Debye screening length, $1/m_D$, or the inter-ion distance $n^{-1/3}$. As such, it can  add coherently over a large number of ions and dominate over the strong but short-ranged Yukawa term in determining the crystal structure. If that is the case, the ions will be spaced by the distance $a_\parallel = \pi/k_e$ (or a multiple of $\pi/k_e$) in the longitudinal direction in order to be at the bottom of the $\cos(2 k_e z)$ oscillations (a scenario already suggested in \cite{2011PhRvC..83b5803S}) . To verify this prediction and determine the crystal structure in the direction perpendicular to the magnetic field, one would, in principle, have to perform a calculation of the free energy of different crystal structures, including  thermal effects. This is particularly difficult given the long range of the forces involved which requires numerical calculations to  consider systems of many tens of thousands of ions. We will perform a much simpler calculation and argue that it captures the essential features of the problem. More specifically, there are two ways that temperature effects enter into our problem. The first is in the motion of the electrons and consequently on the screened potential between ions.  These effects, due to the smearing of the Fermi surface at finite temperature,  are fully taken into account by using the temperature dependent value of $\lambda_T=2\pi k_e/(mT)$ \citep{1971qtmp.book.....F} in Equation~(\ref{eq:V_ion}). As the temperature decreases, $\lambda_T$ increases and the coherent sum  of the Friedel oscillations contributions becomes larger and eventually dominates the energetics of the system. We will neglect, however, the direct effect of the temperature on the motion of ions and minimize the energy of the ion lattice (as opposed to its free energy). This procedure will be justified self-consistently after we discuss the results.

\section{Crystal structure}

The energy of a crystal in the presence of a magnetic field depends on its orientation relative to the field; this is not a small effect for the parameter space we are interested in. Consequently, the search for the most energetically favored arrangements of ions in a lattice has to include not only the usual Bravais lattices but also different orientations of these lattices. We restricted our search to the most likely lattices and to orientations  with a high degree of symmetry in relation to the magnetic field, namely 
simple cubic, face-centered cubic (fcc), and body-centered cubic (bcc) with their four-fold rotation axis along $\mathbf{B}$, hexagonal close-packed (hcp) with its six-fold rotation axis along $\mathbf{B}$, and rotated-bcc with the diagonal axis along $\mathbf{B}$ (in this case $\mathbf{B}$ is fixed to be along the direction from a lattice site to one of its nearest neighbors, ${\bf n}=(1, 1,1)/\sqrt{3}$).

For each crystal structure and for each value of $B$, $T$ and $\rho$ we numerically compute the energies as $a_\parallel$, the lattice constant along the $z$ direction,  varies  over  the range $0 < a_\parallel < 5\pi/(2 k_e)$. The spacing on the transverse direction varies with $a_\parallel$  in order to  keep the density, $\rho$, fixed. This computation requires extra care as the contribution from distant ions can be important.
The method of calculation takes advantage of the fact that the crystals considered have translational symmetry in that the structures appear the same to any two points chosen from the lattice. Hence the energy per particle $E/N$ in such configurations is equal to the total energy for one particle divided by two:
\begin{equation}
\frac{E}{N} = \frac{1}{2} \sum_{i\neq 0}   V(r_{\perp  i}, z_i),
\end{equation} where $i$ indexes the ions in the lattice.
We computed $E/N$ by explicitly adding the contribution of several thousand ions in the longitudinal direction and a few on the transverse directions. The contribution from the remaining ions was estimated by approximating the sum over ions by an integral and we verified that including more ions in the sum (as opposed to the integral) has a negligible impact on the result. 
Since the Friedel potential is cylindrically symmetric, our code traverses the lattices by layers in the z-direction. Lattice symmetries such as mirror symmetry in the x-y plane and rotational symmetries are used to speed up computations. The iteration continues until the additional energy provided by another particle falls below a cut-off threshold. For the results in this paper, a cut-off of $10^{-7}$ was used as it provides a balance between speed and precision. 
After the value of $a_\parallel$ that minimizes the energy is determined, a similar calculation is performed for the other lattices/orientations discussed above. A Friedel crystal is said to form when $a_\parallel$ is equal to an integer times $\pi/k_e$. The result of this calculation for a fixed value of $B$, $T$ and $\rho$ is represented by one point in Fig.~\ref{fig:crystals}. The color coding in Fig.~\ref{fig:crystals} is as follows: red means fcc is the favored crystal structure, blue is bcc, green is rotated-bcc and yellow is hcp. Fig.~\ref{fig:crystals} shows the results for the range  $10^5$~g/cm$^3< \rho <  10^{10}$ g/cm$^3$, $10^{13}$ G$ < B < 10^{16}$ G and 0.1\ keV $< T < $10\ keV ( $1.16\times10^6$ K $< T  < 1.16\times 10^8$ K). As shown in this plot,  Friedel crystals are favored  in the low temperature regime. This is expected since a smaller value of $T$ implies in a longer range for the Friedel part of the potential and consequently a larger coherent sum over the longitudinal direction that drives the ions to be separated by a distance close to $a_\parallel\approx \pi/k_e$.  It is important to stress that the absence of  Friedel crystal in a certain region in thr  parameter space does not mean that the material is not solid. On the contrary, standard Coulomb bcc or hcp crystals are expected in most of these regions \citep{1982JPSJ...51.3431N,1983JPSJ...52...44N,2009PhRvE..80d6405B}. These crystals are not, however, dominated by the mechanism due to the Friedel oscillations that is the subject of this paper and the temperature at which they melt has been previously studied  \citep{1993PhRvE..47.4330F,PhysRevE.62.8554}.
Another feature worth mentioning is that in going from the low densities to the high densities at a fixed value of temperature and magnetic field we find that the favored crystal structure changes from fcc to hcp and then to rotated-bcc and again to fcc. We have no simple intuitive explanation of this fact as it comes about through the delicate competition of many features. We should note that at high densities the energy difference for rotated-bcc and fcc crystals are very small and they are essentially degenerate. 

\begin{figure}
\includegraphics[scale=0.35]{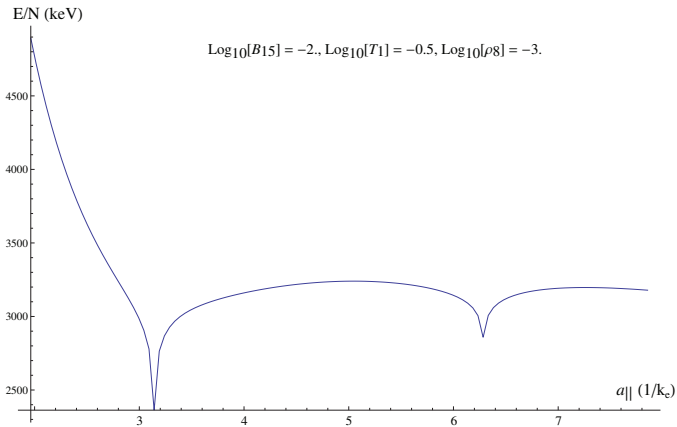}%
\caption{\label{fig:E_a} Energy per ion as a function of $a_\parallel$ for $B=3\times 10^{13}$ G, $\rho=10^5$ g/cm$^3$ and $T\approx 3.7\times10^6$ K. It is shown that the depth of the global minimum at $a_\parallel=\frac{\pi}{k_e}$ is much larger than the temperature and therefore neglecting the effect of thermal fluctuations on the lattice structure determination is not a bad assumption.}
\end{figure}

We now discuss the validity of neglecting the effect of thermal fluctuations on the ions position in the lattice structure determination.
A popular criterion used to estimate the melting temperature of a Coulomb crystal consists of equating the potential energy per ion to the temperature resulting in
\begin{equation}\label{eq:criterion}
T_{melt} \sim \Gamma \frac{Z^2 \alpha}{l},
\end{equation} where $l$ is the typical interparticle distance and the constant $\Gamma \approx 175$ can be computed by Monte Carlo simulations \citep{1993PhRvE..47.4330F,PhysRevLett.104.231101}. This criterion is not valid for our case where the energetics of the system are dominated by long-range forces. An analogous criterion, however, can be obtained by estimating the potential energy per particle by the sum of the Friedel term of the potential along the longitudinal direction. The result is
\begin{equation}
T_{melt} \sim \frac{Z^2 \alpha m_D^2}{k_e} \ln(k_e \lambda_T).
\end{equation} This criterion suggests that temperature effects will be negligible for all temperatures below  $1\ MeV\left ( A_{66}^2 B_{15}^3/\rho_8^2\right )$.
 An alternative way of thinking about this criterion to estimate the importance of these effects is the comparison between the distance $a_\parallel$ and the fluctuations  of $a_\parallel$. As long as the fluctuations on $a_\parallel$ are significantly smaller than $a_\parallel$ we can be assured that the minimization of the potential energy we performed is legitimate. 
Figure~\ref{fig:E_a} shows a typical example of the dependence of the potential energy per ion, $E/N$, as a function of $a_\parallel$. Its pronounced minimum at $a_a\approx \pi/k_e$ has the logarithmic shape expected from the sum  along the longitudinal direction  of $V(r_\perp=0,z) \sim 1/z$ with a cutoff at $ z \approx \lambda_T$  due to the $e^{-z/\lambda_T}$ in Equation~(\ref{eq:V_ion}),  and it is deeper at smaller $T$. In order for the fluctuations to destroy the crystal they have to be large enough to ``climb up'' the potential well. 
The thermal fluctuations were previously estimated analytically  in \citet{2013PhRvC..88e5801B} in a simple model using a quadratic approximation for the bottom of the potential well and it is equivalent to Equation~(\ref{eq:criterion}). We found that the quadratic approximation is not a good approximation for the shape of the $E/N$ vs. $a_\parallel$ curves we computed for some regions of the parameter space. Thus we explicitly verified for every value of the parameters $B$, $T$ and $\rho$ that the height of the potential well was larger than the temperature $T$ by a factor of 2 or (usually much) more.  
A better and even more definitive determination of the effect of non-zero temperature on the properties of the crystal can only be achieved through Monte Carlo calculations.

\section{Summary}

In this paper we study the crystal structure of ions at the low densities and high magnetic fields relevant to the outer crust of magnetars. We found that ions in some parts of our parameter space form a novel crystalline phase made of strongly coupled filaments along the magnetic field, where the spacing between ions in the longitudinal direction is fixed by a multiple of $\pi/k_e$, and more loosely bound in the transverse direction (called Friedel crystals). The underlying cause of this is the long-range oscillatory behavior in the ion-ion potential along the magnetic field due to the anisotropic electronic screening of ions in this direction \citep{2011PhRvC..83b5803S}, an effect that has been neglected in  previous studies of the magnetar crust structure. In this paper we carry out an energy minimization for different types of crystals (simple cubic, fcc, bcc, rotated-bcc and hcp) to determine the exact structure of the Friedel crystals. 
The regions of the parameter space ($T$, $B$ and $\rho$) where Friedel crystals are formed is in qualitative agreement with the estimates in \cite{2013PhRvC..88e5801B}. 
Our calculations predict the formation of Friedel crystals in the low temperatures ($10^6$ to $5\times 10^7$ K), but even at $T\sim 5\times 10^8$ K, this phase exists for some values of $B$ and $\rho$.
 
\begin{figure*}
\begin{center}
\includegraphics[scale=0.8]{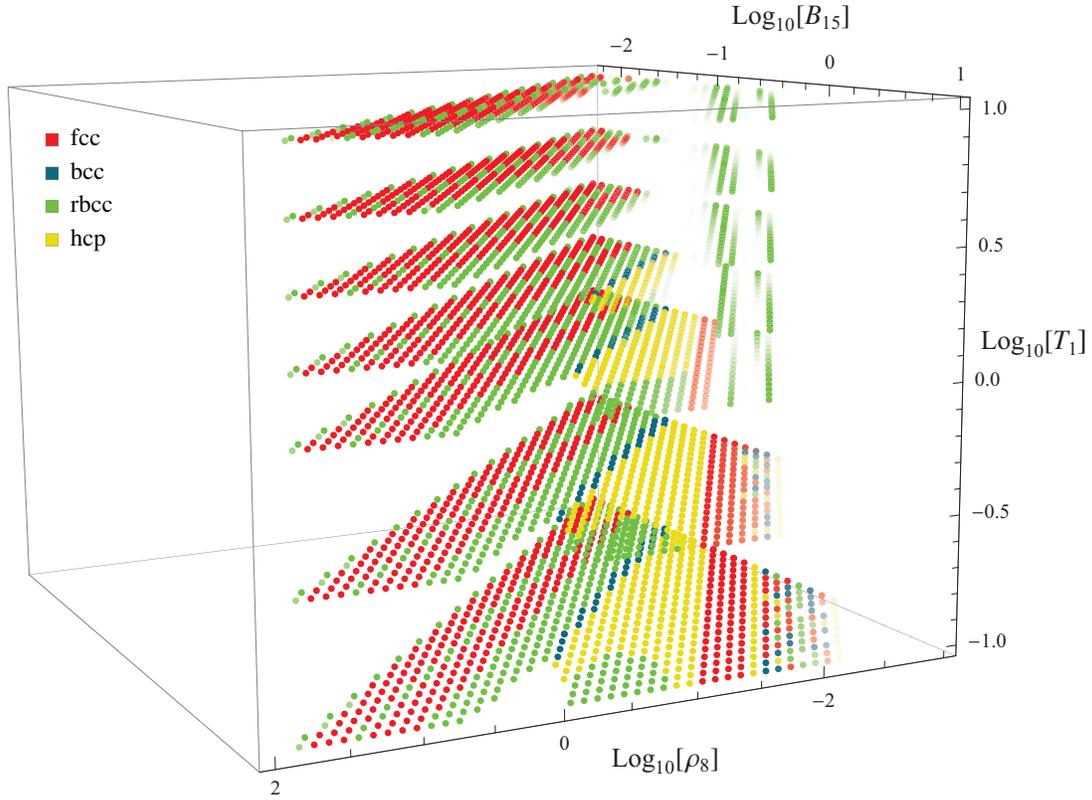}%
\end{center}
\caption{\label{fig:crystals}  The regions where Friedel crystals form in the parameter space relevant to the outer crust of highly magnetized neutron stars ($10^5$~g/cm$^3< \rho <  10^{10}$ g/cm$^3$, $10^{13}$ G$ < B < 10^{16}$ G and 0.1\ keV $< T < $10\ keV ) for fully ionized $^{66}$Ni$_{28}$ matter. Red shows the region where fcc is the favored crystal structure, blue is bcc, green is rotated-bcc and yellow is hcp. We note that the absence of Friedel crystals in some regions of the parameter space does not mean that the material is not solid; in contrast, in most of these regions ions form a standard Coulomb crystal but the structure of these crystals along the magnetic field is not determined by the Friedel oscillations in the ion-ion potential. (A color version of this figure is available in the online journal.)}
\end{figure*}

The present calculation, besides being  more rigorous than the one in  \cite{2013PhRvC..88e5801B}, also determines the crystal structure favored for each value of $T$, $B$ and $\rho$. 
As can be seen in Fig.~\ref{fig:crystals} the phase diagram is rich, but roughly speaking, the favored crystal structures at high temperatures are fcc and rotated-bcc (where the magnetic field is along the diagonal axis), and going to lower temperatures and lower densities at a fixed value of the magnetic field, hcp becomes more favorable. The determination of the crystal structure paves the way to the computation of all elastic moduli, including the ones that depend on the ionic distances in the transverse direction.
This has potentially interesting implications for the magnetar QPOs, since the fundamental shear modes in neutron star crusts are very sensitive to the properties of the crust. The long-range Friedel potential will likely affect the breaking strain of the crust as well which would have implications for gravitational wave physics by allowing for larger deformations of the crust.

\acknowledgments
We thank Ittai Baum for helpful discussions. We acknowledge the support of the U.S. Department of Energy through grant number DEFG02- 93ER-40762.

\bibliography{crust_structure}


\end{document}